\documentclass[fleqn,usenatbib]{mnras}
\usepackage{float}

\usepackage{newtxtext}
\usepackage{color, xcolor}
\usepackage{soul}
\soulregister\cite7  
\soulregister\citep7 
\soulregister\citet7 
\soulregister\ref7 
\soulregister\pageref7 

\usepackage[T1]{fontenc}

\DeclareRobustCommand{\VAN}[3]{#2}
\let\VANthebibliography\thebibliography
\def\thebibliography{\DeclareRobustCommand{\VAN}[3]{##3}\VANthebibliography}

\usepackage{graphicx}	
\usepackage{amsmath}	
\usepackage{amssymb}	

\title[TDE outflows and PeV neutrino]{Could TDE outflows produce the PeV neutrino events?}

\author[Wu et al.]{
	Han-Ji Wu,$^{1,2}$
	Guobin Mou,$^{1,2}$\thanks{gbmou@whu.edu.cn}
	Kai Wang,$^{3}$\thanks{kaiwang@hust.edu.cn}
	Wei Wang,$^{1,2}$\thanks{wangwei2017@whu.edu.cn}
    Zhuo Li$^{4,5}$\thanks{zhuo.li@pku.edu.cn}
	\\
	$^{1}$School of Physics and Technology, Wuhan University, Wuhan 430072, China \\
	$^{2}$WHU-NAOC Joint Center for Astronomy, Wuhan University, Wuhan 430072, China \\
	$^{3}$Department of Astronomy, School of Physics, Huazhong University of Science and Technology, Wuhan 430074,  China  \\
    $^{4}$Department of Astronomy, School of Physics, Peking University, Beijing 100871, China\\
    $^{5}$Kavli Institute for Astronomy and Astrophysics, Peking University, Beijing 100871, China
}

\date{Accepted XXX. Received YYY; in original form ZZZ}

\pubyear{2021}

\begin{document}
	\label{firstpage}
	\pagerange{\pageref{firstpage}--\pageref{lastpage}}
	\maketitle
	
\begin{abstract}
	A tidal disruption event (TDE), AT2019dsg, was observed to be associated with a PeV neutrino event, IceCube-191001A, lagging the optical outburst by a half year. It is known that TDEs may generate ultrafast outflows. If the TDE occurs in a cloudy environment, the outflow-cloud interactions may form shock waves which generate accelerated protons and hence delayed neutrinos from hadronic interactions in clouds. Here we investigate the neutrino production in AT2019dsg by examining the TDE outflow-cloud interaction model. We find  that, for an outflow with a velocity of 0.07c and a kinetic luminosity of $10^{45}\rm erg\ s^{-1}$, protons may be accelerated up to $\sim$ 60 PeV by the bow shocks, and generate PeV neutrinos by interactions with clouds. 
 The predicted neutrino number in this model depends on the uncertainties of model parameters and in order to match the observations, some challenging values of parameters have been involved. The PeV neutrino event number can be $\sim 4\times10^{-3}$ for a hard proton index $\Gamma=1.5$.   
	\end{abstract}
	
	
	
    \section{Introduction}


 On 22 September 2017, IceCube reported a neutrino event with energy $\sim 290\ \rm TeV$, which was shown to be associated with the blazar TXS 0506+056 \citep{aartsen1807science}. This opened a window of high energy neutrino astrophysics. The origin of high energy neutrinos is not clear yet, and TDEs are the other possible sources. 
 
 Recently, \cite{stein2021tidal} reported a correlation between a neutrino event with energy $\sim$0.2 PeV detected by IceCube on 1 October 2019 (IceCube-191001A) and a TDE (AT2019dsg) discovered by Zwicky Transient Facility (ZTF), and the neutrino event lags the onset of TDE by 6 months. AT2019dsg's redshift is $z=0.051$, i.e., the luminosity distance is $D=230$ Mpc. 
 The optical luminosity was observed to decrease from $10^{44.5} \rm erg\ s^{-1}$ to $10^{43.5} \rm erg\ s^{-1}$ \citep{van2020} on a timescale of a half year. AT2019dsg is among the top 10\% of the 40 known optical TDEs in luminosities. The peak radiation is well described by a  $10^{14}\rm cm$-sized blackbody photosphere of temperature $10^{4.59}$ K. 
 AT2019dsg is an unusual TDE \citep{2102.11879}; it belongs to neither the typical soft-X-ray TDEs nor the typical optical/UV TDEs because it emits optical/UV radiations as well as X-ray and radio radiations. 
 Fermi Large Area Telescope (Fermi-LAT) provides an upper limit of flux $1.2\times 10^{-11}\rm erg\ cm^{-2}\ s^{-1}$ in 0.1-800 GeV range. HAWC observatory also set an upper limit for the period from 30 September to 2 October, $ E^{2}\Phi=3.51\times 10^{-13}(\frac{E}{\rm TeV})^{-0.3}\rm TeV\ cm^{-2}\ s^{-1} $ for 300 GeV-100 TeV \citep{van2020}.
    

 In the previous literatures (e.g., \citealt{wang2016tidal,liu2020}), a jet from the TDE is assumed to accelerate protons and generate neutrinos via p$\gamma$ interactions, dominating over pp interactions since the density of photons is extremely high. Moreover, in the TDE model by \cite{murase2020high} high-energy neutrinos and soft gamma-rays may be produced via hadronic interactions in the corona, the radiatively inefficient accretion flows and the hidden sub-relativistic wind. 


  It is known that, in addition to the radiative outburst, TDE is also expected to launch ultra-fast and energetic outflows. First, due to the  relativistic apsidal precession, after passing the pericenter, the stream collides with the still falling debris, leading to a collision-induced outflow \citep{lu2020self}. Second, after debris settling into an accretion disk, the high accretion mode will launch energetic outflows \citep{curd2019grrmhd}. Since the duration of both processes above are of the order of months, the duration of launching energetic TDE outflows is also roughly in months. For AT2019dsg, the physics of the outflow is estimated via radio emission. The velocity inferred in different models are similar: 0.07 c in outflow--circumnuclear medium (CNM) interaction model \cite{cendes2021radio}, or 0.06--0.1c in outflow--cloud interaction model \citep{mou2021radio}). However, the outflow energy or kinetic luminosity of outflows ranges a lot in different models. \citet{cendes2021radio} estimated an energy of $4\times 10^{48}$ erg, which is much smaller than the energy budget of the TDE system ($\sim 10^{53}$ erg).  \citet{mou2021radio} inferred that the outflow power should be around $10^{44}$ erg s$^{-1}$ which is consistent with numerical simulations \citep{curd2019grrmhd}, and the total energy should be in the order of $10^{51}$ erg if the outflow continues for months.  
  
    
 When the outflow impacts a cloud, the bow shock (BS) can be produced outside the cloud, and could effectively accelerate particles via diffusive shock acceleration (DSA) processes. 
 The electrons accelerated at BSs give rise to synchrotron emission, which can be detected in $\sim$GHz radio band \citep{mou2021radio}. The accelerated high energy protons may be the precursor of the neutrinos, especially considering the high-density cloud near the BS  which is favorable for pp collisions \citep{mou2021years2}. A basic premise is whether there are clouds around the black hole, especially at the distance of $10^{-2}$ pc as inferred from the delay of the neutrino event and the possible outflow speed. It is well known that for active galactic nuclei (AGN), there exists a so-called broad line region (BLR) around the supper massive black hole (SMBH) \citep{antonucci1993unified}, which is frequently referred to as ``clouds''. However, for quiescent SMBH or low-luminosity AGN (the case for AT2019dsg), due to the lack of ionizing radiation irradiating the clouds, the existence of the clouds becomes difficult to verify, and the physics of such clouds remains largely unknown. 
 
 To distinguish from the BLR cloud concept in AGN, we hereby call the possibly existing clouds around quiescent or low-luminosity SMBH at similar position of the BLR as ``dark clouds''. Transient events may help reveal the physics of dark clouds. For AT2019dsg, in addition to the indirect speculation of the existence of dark clouds via radio emission \citep{mou2021radio}, direct evidences arise from the dusty echo, and broad emission line components. 
 First, \cite{van2021establishing} reported that AT2019dsg was detected with remarkable infrared echo about 30 days after the optical onset, suggesting that clouds should exist at a distance of 0.02 pc to the SMBH (note that there may exist clouds in more inward regions but not surveyed by WISE/neoWISE). Second, it is reported that there exist the broad emission line components (line widths $>$ 70 \r{A}) of H$\alpha$, H$\beta$, and He\uppercase\expandafter{\romannumeral2} \citep{cannizzaro2021accretion}, implying the existence of materials of the velocity over several thousand kilometers per second, although their nature is unclear.
 In our TDE outflow-cloud interaction model, we assume that there exist dark clouds at $\sim 0.01$ pc to the BH, and we simply set up the parameters (covering factor, cloud size and density) of dark clouds similar referring to those of classical BLR clouds.
	
	
 This paper is organized as follows. We introduce the general physics picture of the model in Sec.~\ref{Pp}; The products (GeV-TeV gamma-rays and PeV neutrinos) from hadronic emission are described in \sethlcolor{red}Sec.~\ref{dp}; In Sec.~\ref{cwo}, we compare the calculations with the present observations; The conclusions and discussion are presented in the last section.

  \section{Physical picture of outflow-cloud interactions}\label{Pp}
  
\begin{figure*}
		\centering
        \includegraphics[scale=0.8]{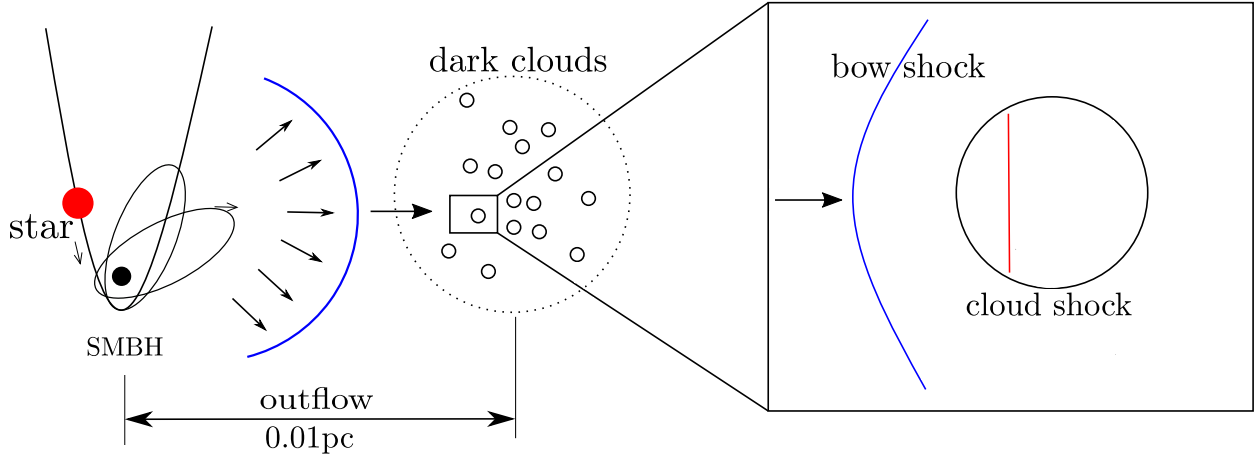}
		\caption{Schematic plot of the TDE outlfow--cloud interaction model. A star engages the Roche radius of the SMBH thereby the disrupted debris is blown away. The collision-induced outflows hit clouds forming shock waves, a bow shock and a cloud shock. 
		The outflow--cloud interactions occur at 0.01 pc from the SMBH. 
		See section \ref{poo} for details. }
		\label{S}
	\end{figure*}

As shown in Fig. \ref{S}, consider that there are dark clouds surrounding the SMBH.
The TDE outflows released from the SMBH collide with the dark clouds, forming two shock waves \citep{mckee1975}, i.e., a bow shock (BS) outside the cloud and a cloud shock (CS) sweeping through the cloud. Following \cite{celli2020spectral}, protons may be accelerated to very high energy with a power-law distribution. The high-energy protons may partly propagate into the cloud and interact with matter therein.

	\subsection{Dynamics}\label{poo}
	
We consider a simplified spherically symmetric outflow, with kinetic luminosity $L_{\rm kin}$ and velocity $V_{\rm o}$. We take $L_{\rm kin}=10^{45} \rm erg~s^{-1}$ as the fiducial value \citep{curd2019grrmhd}, which is also close to the constraint given by \cite{mou2021radio}. Following the interpretation of the radio observation \citep{stein2021tidal} by synchrotron radiation from non-thermal electrons in the outflow-CNM model, we take the outflow velocity derived from the model, $V_{\rm o} = 0.07\rm c$ \citep{cendes2021radio}. The duration of the outflow launching is assumed to be $T_{\rm o}\sim 6$ months.

Define $\rho_{\rm o}$ the density of outflow material, then we write the kinetic luminosity as $L_{\rm kin}=\frac{1}{2}\dot{M_{\rm o}}V_{\rm o}^{2}=2\pi r_{\rm o}^{2}\rho_{\rm o}V_{\rm o}^3$, with $\dot{M_{\rm o}}$ the mass flowing rate. Since the time delay of the neutrino event and the TDE is $t_{\rm delay}\sim 6$ month \citep{van2020}, we assume the typical distance of the dark clouds from the central SMBH is $r_{\rm o}=V_{\rm o}t_{\rm delay} \simeq 0.01$ pc. 
 Thus the number density of the outflow is $n_{\rm o}=\frac{\rho_{\rm o}}{m_{\rm H}}\sim 1.14\times 10^{7}(\frac{L_{\rm kin}}{{{10^{45}\rm erg\ s^{-1}}}})(\frac{V_{\rm o}}{0.07{\rm c}})^{-3}(\frac{r_{\rm o}}{0.01{\rm pc}})^{-2}\rm cm^{-3}$.

The interaction between outflows and clouds drives a CS that sweeps across the cloud. The velocity of CS is related to the outflow velocity by $V_{\rm c}=\chi^{-0.5}V_{\rm o}$ \citep{mckee1975, mou2020years1}, where $\chi\equiv \frac{n_{\rm c}}{n_{\rm o}}$, with $n_{\rm c}$ the particle density in the cloud. According to the photoionization models in BLRs of AGN, we assume the cloud particle density to be $n_{\rm c} \sim 10^{10} \rm cm^{-3}$ \citep{osterbrock2006astrophysics}. So, here $\chi\simeq 8.8\times 10^{2}$.

Let us assume the size of the clouds is typically $r_{\rm c}=10^{14}$cm\footnote{This can be obtained from the column density ($\sim10^{24}{\rm cm^{-2}}$) and the cloud density $ n_{\rm c}\sim10^{10}\rm cm^{-3}$ \citep{osterbrock2006astrophysics}.}. The CS crosses the cloud in a timescale of 
\begin{equation}
T_{\rm cloud}=\frac{r_{\rm c}}{V_{\rm c}}=\frac{r_{\rm c}}{V_{\rm o}} \chi^{0.5},
\end{equation}
i.e., $T_{\rm cloud}\sim1(\frac{r_{\rm c}}{10^{14}\rm cm})(\frac{V_{\rm o}}{0.07\rm c})^{-1}(\frac{n_{\rm c}}{10^{10}\rm cm^{-3}})^{0.5}(\frac{n_{\rm o}}{1.14 \times 10^{7}\rm cm^{-3}})^{-0.5}$ month. Note, the cloud could be devastated by the outflow, and the survival timescale of the clouds after the CS crossing is comparable to $T_{\rm cloud}$ \citep{klein1994hydrodynamic}, so $T_{\rm cloud}$ can also be regarded as the survival timescale of the cloud.

	\subsection{Particle acceleration and propagation}
	\label{swa}
 Both BS and CS may accelerate particles. According to DSA mechanism, the acceleration timescale in the BS for a particle with energy $E_p$ and charge number $Z$ is \citep{drury1983introduction}
	\begin{equation}
		T_{\rm acc,BS}\approx\frac{8}{3}\frac{cE_{\rm p}}{ZeB_{\rm o}V_{\rm o}^2},
		\label{AM}
	\end{equation}
    where  $B_o$ is the magnetic field strength in the outflow. For the CS, the acceleration timescale is 
    	\begin{equation}
		T_{\rm acc,CS}\approx\frac{8}{3}\frac{cE_{\rm p}}{ZeB_{\rm c}V_{\rm c}^2}
	\end{equation}
	where $B_{\rm c}$ is the magnetic field in the cloud. We will assume $B_{\rm o}= B_{\rm c}=B$ in the nearby region of the outflow--cloud interaction, and we take $B=1$ G.

The particle acceleration suffers from several factors. The first is the particle energy loss due to hadronic interactions, namely the pp interactions. In the BS, the timescale is 
	\begin{equation}
	     t_{\rm pp,BS}=\frac{1}{cn_{\rm o}\sigma_{\rm pp}} ,\label{tpp}
	\end{equation}
whereas in the CS, 
	\begin{equation}
	     t_{\rm pp,CS}=\frac{1}{cn_{\rm c}\sigma_{\rm pp}} .\label{tpp,cs}
	\end{equation}
Here $\sigma_{\rm pp}\simeq 30$mb is the pp cross section. The other suppression factor is the lifetime of the relevant shocks. As the cloud survival timescale is comparable to the CS crossing time, and after the cloud is destroyed,  both BS and CS end. So the acceleration in either BS or CS is only available within a time period of $T_{\rm cloud}$ \citep{klein1994hydrodynamic}. Finally, the maximum energy of the accelerated particles is determined by equating the acceleration time to the shortest one between the pp interaction time and the CS crossing time of the cloud.

\begin{figure}
    	\centering
    	\includegraphics[scale=0.6]{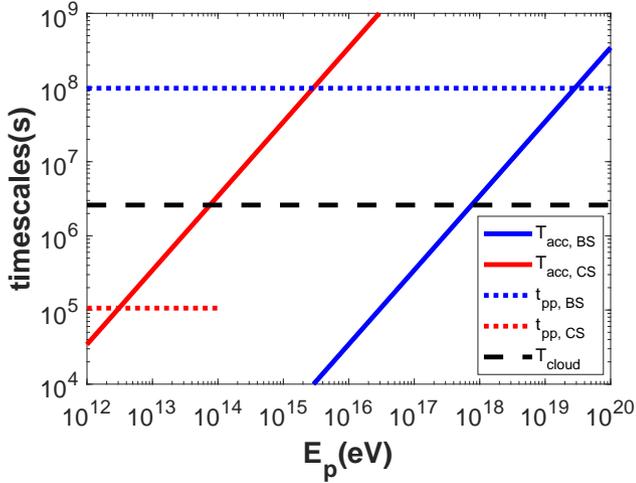}
    	\caption{The timescales of particle acceleration (solid) and pp interaction (dotted) in the BS (blue) and CS (red), and the cloud survival timescale (black). } 
    	\label{tt}
    \end{figure}

All the timescales are plotted in Fig. \ref{tt}.  
For the BS, $T_{\rm cloud} \sim 1$ month is the main constraint than the pp energy loss time, due to the low density in the outflow, $t_{\rm pp,BS}=3.1(\frac{n_{\rm o}}{1.14\times 10^{7}{\rm cm^{-3}}})^{-1}$yr. By equating $T_{\rm acc,BS}=T_{\rm cloud}$ we obtain the maximum energy of particles accelerated in the BS, $E_{\rm p,max}\simeq 60(\frac{B}{1{\rm G}})(\frac{V_{\rm o}}{0.07{\rm c}})^{2}(\frac{T_{\rm acc,BS}}{{\rm 1\ month}})$PeV.
For the CS, due to the dense cloud, the pp collision time is short, $t_{\rm pp,CS}=1(\frac{n_{\rm c}}{10^{10}\rm cm^{-3}})^{-1}$ day, and is more important in suppressing acceleration. By $T_{\rm acc,CS}=t_{\rm pp,CS}$, one obtains the maximum energy $E_{\rm p,max} \simeq 2.9(\frac{B}{1{\rm G}})(\frac{V_{\rm c}}{7.1\times 10^{7}{\rm cm\ s^{-1}}})^{2}(\frac{T_{\rm acc,CS}}{1 {\rm day}})$TeV. Thus only the BS can accelerate particles up to PeV scale.

 
Note that we neglect the effect of energy loss due to p$\gamma$ interactions between the high-energy particles and the TDE photons. Given the cross section $\sigma_{\rm p\gamma}\sim0.2$mb on average, and the TDE photon number density at $r_{\rm o}$, $n_{\rm ph}\simeq 10^{9} \rm cm^{-3}$ (see Sec.~\ref{gammarays}), the timescale of $\rm p\gamma$ interactions is relatively large, $t_{\rm p\gamma}\sim 3.2(\frac{n_{\rm ph}}{1\times 10^{9}{\rm cm^{-3}}})^{-1}\rm yr$. In the previous works (e.g., \citealt{liu2020}), $\rm p\gamma$ reaction is important because a site closer to the center is considered, so the photon density is high, $n_{\rm ph}\sim 10^{16}(\frac{L}{10^{43}\rm erg\ s^{-1}})(\frac{r_{\rm o}}{10^{14.5}\rm cm})^{-2}\rm cm^{-3}$ (with $L$ being the TDE luminosity, see below) and $\rm p\gamma$ reaction is important. In our case, neither p$\gamma$ nor pp reactions in the BS consume significant energy of the accelerated particles. 
     
     After acceleration in the BS, the high-energy particles may diffuse away from the BS \citep{bultinck2010origin,taylor1961diffusion,kaufman1990explanation}. As suggested by some literatures \citep{1997ApJ...478L...5D,2010ApJ...724.1517B,2012ApJ...755..170B,2018arXiv180900601W}, we assume that a significant fraction $F$ of the accelerated particles can effectively reach and enter the cloud, whereas the other propagate away bypassing the cloud. Basically, the relatively low-energy protons are likely to be advected with the shocked material, 
	while the relatively energetic particles ($\gtrsim  1\,\rm TeV$) tend to diffuse up to the cloud. Besides, these high-energy particles entering the cloud could be more important by suppressing the possible advection escape under the certain magnetic configuration  \citep{bosch2012}. The detailed treatment of the particle propagation is beyond the scope of this paper. To evaluate this uncertainty, $F\simeq 0.5$ is invoked in our calculations. For the other high-energy particles that do not propagate into the cloud, no hadronic interactions are expected, given the low density of cold particles outside the cloud.
	
	After entering the cloud, the particles may propagate in the cloud by diffusion. The residence time in the cloud before escaping can be estimated by
    \begin{equation}
        \tau_{\rm es}=C_{\rm e}\frac{r_{\rm c}^{2}}{D_{\rm B}},
        \label{eqescape}
    \end{equation}
    where $D_{\rm B}$ is the Bohm diffusion coefficient. and $C_{\rm e}$ is a correction factor that accounts for the difference between the actual diffusion coefficient and the Bohm diffusion. We take $C_{\rm e}=0.75$. The Bohm diffusion coefficient of protons is given by \citep{kaufman1990explanation} $D_{\rm B}=\frac{r_{\rm g}^{2}\omega_{\rm g}}{16}$, where $r_{\rm g}=\frac{E_{\rm p}}{eB}$ is the cyclotron radius, and $\omega_{\rm g}= \frac{eBc}{E_{\rm p}}$ is the cyclotron frequency. Thus we get $\tau_{\rm es}\sim 1.3(\frac{E_{\rm p}}{7\rm PeV})^{-1}(\frac{B}{1\rm G})$day, a value similar to $t_{\rm pp,CS}$ for $E_{\rm p}\sim7$ PeV. 
    

    \section{Hadronic emission}\label{dp}
    
   In the outflow--cloud interaction, the kinetic energy of the outflow will be converted into the BS and CS. The energy ratio between the CS and BS is $\chi^{-0.5}\simeq0.034$, so the energy dissipation in the CS can be neglected compared to the BS (see appendix A in \citealt{mou2020years1}).
   
The covering factor of the clouds is $C_{\rm v}\sim 0.1$, and the shock acceleration efficiency, i.e., the fraction of the shock energy converted to accelerated particles, is $\eta \sim 0.1$. Given the kinetic energy of the outflow, the average luminosity of the accelerated particles is, in the BS,
    \begin{equation}
    	L_{\rm b}=C_{\rm v}\eta L_{\rm kin}, \label{Eb}
    \end{equation}
    and in the CS,
    \begin{equation}
    L_{\rm c}=C_{\rm v}\chi^{-0.5}\eta L_{\rm kin}. \label{Eb}
    \end{equation}
  Plugging the numbers, $L_{\rm c}\approx 3.4\times 10^{41}\rm erg\ s^{-1}$, and $L_{\rm b}\approx  10^{43}\rm erg\ s^{-1}$ for $\eta=0.1$. The luminosity of CSs is so small, which allow us to neglect their contribution in emission.
    
    We assume the accelerated relativistic particles distribute as a power-law spectrum with spectral index $\Gamma$ and an exponential cut-off at the high energy:
    \begin{equation}
    	\frac{dn(E_{\rm p})}{dE_{\rm p}}=K_{\rm p}E_{\rm p}^{-\Gamma}e^{-\frac{E_{\rm p}}{E_{\rm p,max}}}\label{Ep}.
    \end{equation}
    The normalization factor $K_{\rm p}$ can be determined by the normalization of the particle luminosity $L_{\rm p}=\int E_{\rm p}\frac{dn(E_{\rm p})}{dE_{\rm p}}dE_{\rm p}$.
   Since neglecting contribution from the CS, we have $L_{\rm p}=L_{\rm b}$. We will consider a range of $\Gamma$ value, from 2 to 1.5 \citep{celli2020spectral}.

The pp collision produce neutral and charged pions,
	\begin{eqnarray}
	&p+p\to p+p+a\pi^{0}+b(\pi^{+}+\pi^{-}),\\
	 &        	p+p\to p+n+\pi^{+}+a\pi^{0}+b(\pi^{+}+\pi^{-}),\label{pp2}
	\end{eqnarray}
	where $a\approx b$. 
	The pions decay and generate $\gamma$-rays and leptons:
	\begin{eqnarray}
	 &		\pi^{0}\to 2\gamma\\
    &	\pi^{+}\to \mu^{+}+\nu_{\mu},~ \mu^{+}\to e^{+}+\nu_{e}+\bar{\nu}_{\mu},\\
    &	\pi^{-}\to \mu^{-}+\bar{\nu}_{\mu},~ \mu^{-}\to e^{-}+\bar{\nu}_{e}+\nu_{\mu}. \label{n2}   
	\end{eqnarray}

	The final product particles produced by pp collisions in the clouds per unit time on average can be given by \citep[e.g.][]{kamae2006,aartsen2014}:
	\begin{equation}
		\frac{dn_{\rm f}}{dE_{\rm f}}= 1.5Fcn_{\rm H} \int\frac{d\sigma(E_{\rm p},E_{\rm f})}{dE_{\rm f}}\frac{dn(E_{\rm p})}{dE_{\rm p}}dE_{\rm p},\label{nf}
	\end{equation}
	where f$=\gamma, \nu$, etc., represents the type of final particles, $\sigma$ is the inclusive cross section as function of final particles and proton's energy, and $n_{\rm H}$ is the background number density of protons.
	The coefficient 1.5 is a correction factor accounting for the contribution of Helium (we assume that the Helium abundance in the BS is similar to Galactic cosmic rays \citep{mori1997galactic}).
Here the integration is calculated using the cparamlib package\footnote{https://github.com/niklask/cparamlib}. If the particle escape from the cloud is fast, the calculated spectrum above should be multiplied by a factor $\frac{\tau_{\rm es}(E_{\rm p})}{t_{\rm pp,CS}}$ to take into account the reduction of secondary products by escape.

		\begin{table*}
		\centering
		\caption{Model parameters}
		\label{123}
		\begin{tabular}{lcr} 
			\hline
			Parameters & Descriptions & Fiducial Values  \\
			\hline
			$L_{\rm kin}$ & the kinetic luminosity of outflow & $10^{45}\rm erg\ s^{-1}$ \\
			$V_{\rm o}$ & the velocity of outflow & 0.07c \\
			$T_{\rm o}$ & outflow launching duration & 6 months \\
			$r_{\rm o}$ & the typical distance of clouds from the SMBH & $0.01\,\rm pc$ \\
			$n_{\rm c}$& the particle density of cloud & $10^{10}\,\rm cm^{-3}$ \\
			$r_{\rm c}$ & the typical size of clouds & $10^{14}\,\rm cm$ \\
			$B$ & magnetic field strength around outflow-cloud interaction region & 1G \\
		    $C_{\rm v}$ & covering factor of clouds & 0.1 \\
		    $\eta$ & the fraction of shock energy converted to accelerated particles & 0.1 \\
			$F$&the fraction of accelerated particles propagate into the cloud  & 0.5\\
			$C_{\rm e}$ & the correction factor of diffusion coefficient relative to Bohm limit & 0.75\\

			\hline
		\end{tabular}
	\end{table*}
	
	\subsection{Neutrino}
	
\begin{figure}
    \includegraphics[scale=0.57]{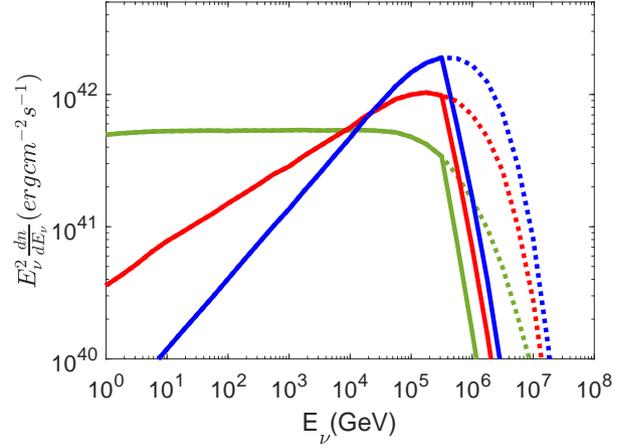}
    \caption{
    The energy distribution of neutrino luminosity. The blue, red and green lines correspond to $\Gamma=1.5$, 1.7, and 2, respectively. The solid (dotted) lines correspond to the cases with (without) the consideration of particle escape from the cloud.  
    }
    \label{NF}
    \end{figure}

	Given the neutrino luminosity and spectrum, we calculate the neutrino event number expected to be detected by IceCube in a time period of $T_{\rm o}$,
	\begin{equation}
		N_{\nu}=\frac{T_{\rm o}}{4\pi D^{2}}\int_{0.1\rm PeV}^{1\rm PeV}dE_{\nu}A_{\rm eff}(E_{\nu})\frac{dn_{\nu}}{dE_{\nu}}.\label{n}
	\end{equation}
   The detected event IceCube-191001A has a neutrino energy of $>0.2$ PeV, thus we only calculate the sub-PeV neutrino events in $0.1-1$ PeV range. The  real time effective area of IceCube is described by \citep{blaufuss2021next}
    \begin{equation}
	    A_{\rm eff}=2.058\times \left(\frac{E_{\nu}}{1\rm TeV}\right)^{0.6611}-32
	\end{equation}
The number of neutrino events is calculated to be $N_{\nu}\simeq 3.5\times10^{-3}$ for $\Gamma=1.5$, considering particle escape from cloud (see details in Fig. \ref{NF}).
    
    
	\subsection{$\gamma$-ray}\label{gammarays}
	
	The intrinsic $\gamma$-ray spectrum accompanying the neutrino emission can also be calculated with equation \ref{nf}, but high-energy $\gamma$-ray photons may be attenuated by interacting with low-energy background photons via $\gamma\gamma\to e^{+}e^{-}$. The background photons may come from the TDE, host galaxy, extragalactic background light (EBL), cosmic microwave background (CMB), and even radiation in the cloud, etc. 
	
	Consider the TDE photons first. At a distance $r_{\rm o}$ from the SMBH, the number density of TDE photons around typical energy $E_{\rm ph}\sim10$eV is estimated by
	\begin{equation}
		n_{\rm ph}=\frac{L}{4\pi r_{\rm o}^{2}cE_{\rm ph}},
		\label{n_ph}
	\end{equation}
	where $L$ is the TDE radiation luminosity, which is given by \cite{stein2021tidal} for AT2019dsg.
    The TDE luminosity may evolve with time, approximately estimated as following the accretion rate evolution, i.e.,
 $	L\propto \dot{M}\propto \left(\frac{t}{T_{\ast}} \right)^{-5/3}$,
    where $T_{\ast}\approx 0.1(\frac{R_{\ast}}{1R_{\odot}})^{3/2}(\frac{M_{\ast}}{1M_{\odot}})^{-1/2}$ yr is the minimum orbit period of the disrupted material \citep{evans1989} depending on the radius $R_{\ast}$ and the mass of the star $M_{\ast}$. The TDE luminosity decreases to about $6\%$ of the peak luminosity after $t_{\rm delay}=6$ months, i.e., $L\sim10^{43}\rm erg\ s^{-1}$. 
    Thus, the TDE photon density is estimated as $n_{\rm ph}\sim 10^{9}(\frac{L}{10^{43}\, \rm erg\ s^{-1}})(\frac{r_{\rm o}}{0.01\,\rm pc})^{-2} (\frac{E_{\rm ph}}{10\,\rm eV})^{-1}\,\rm cm^{-3}$. 
    
\begin{figure}
    	\centering
    	\includegraphics[scale=0.55]{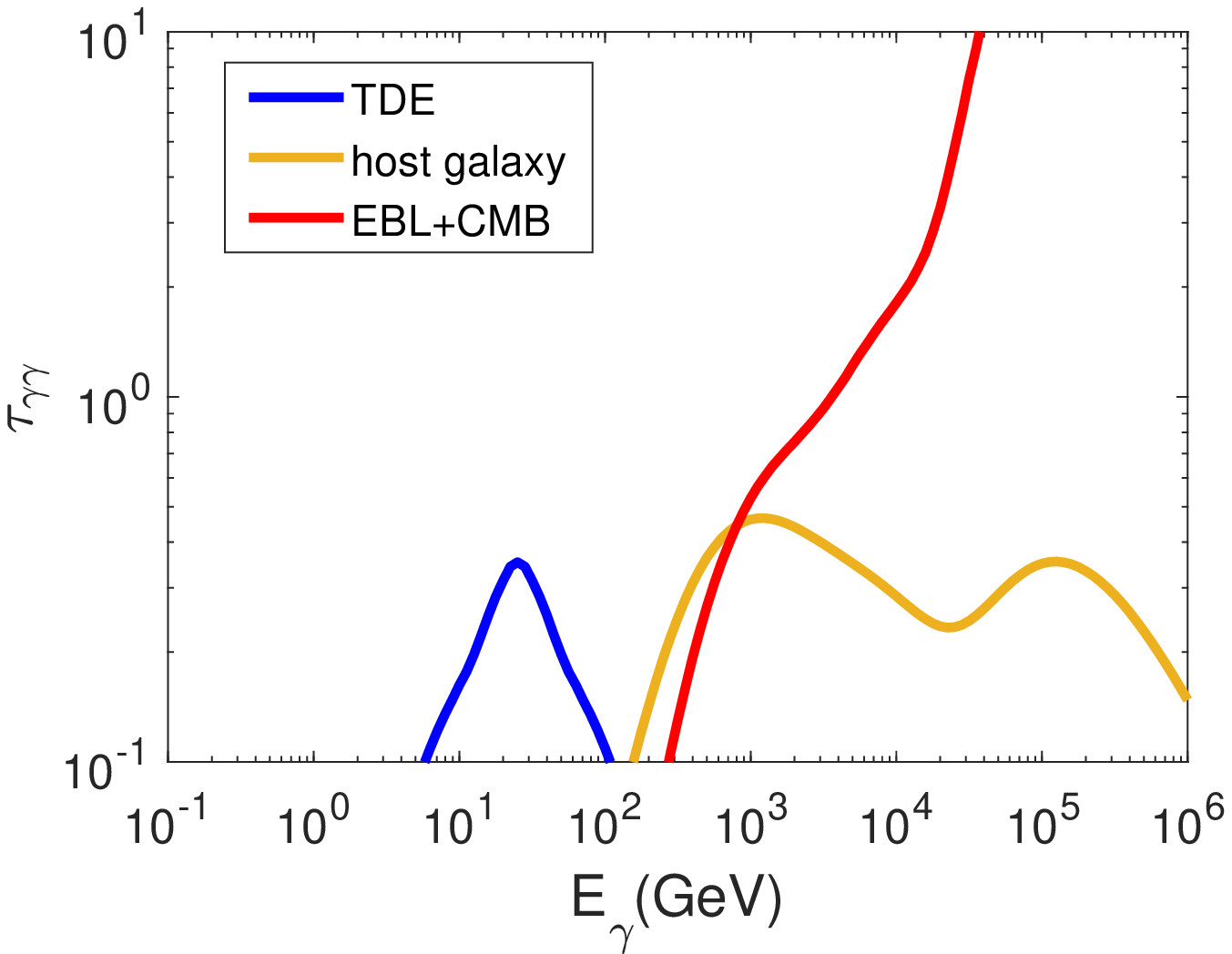}
    	\includegraphics[scale=0.59]{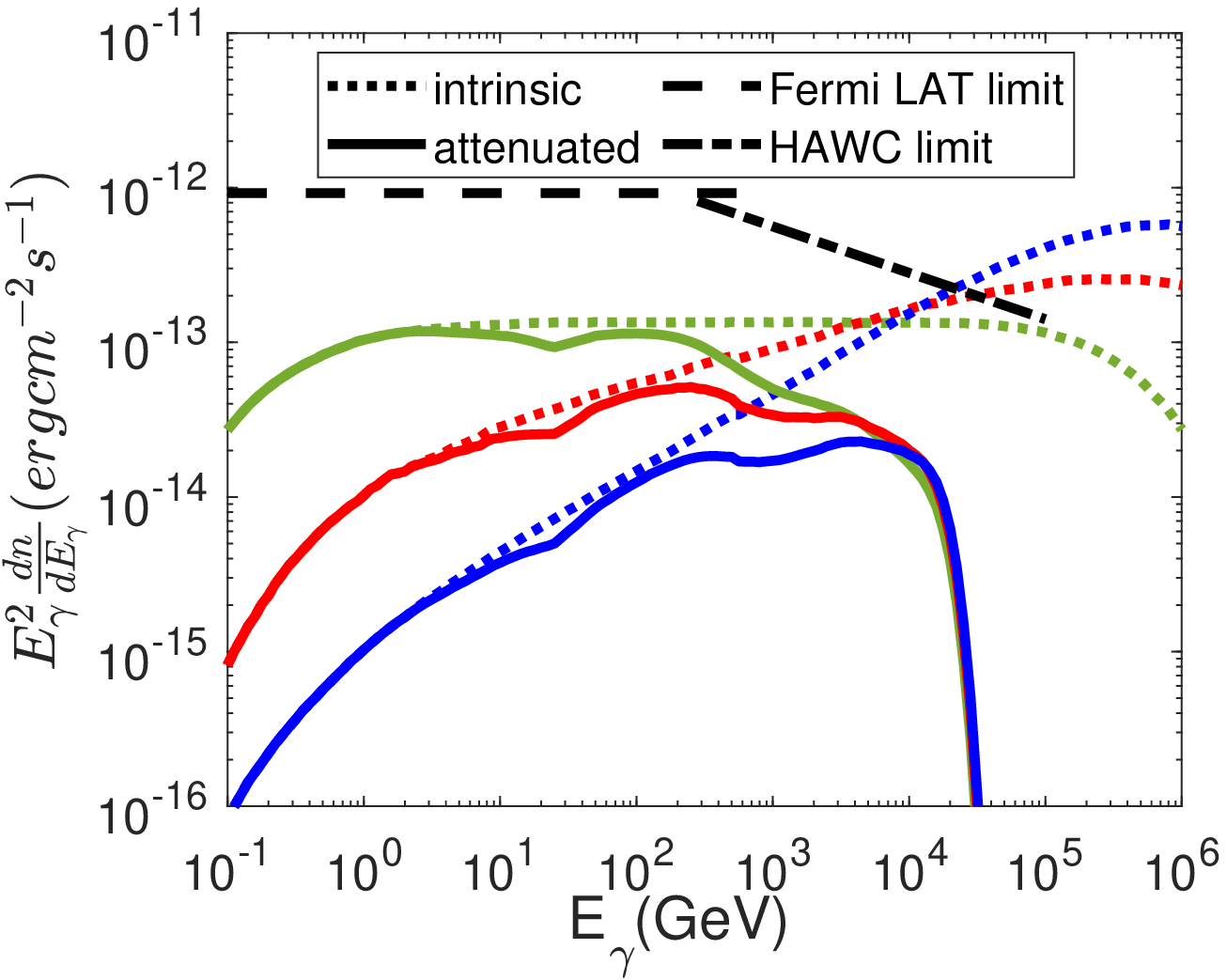}
    	\caption{ The $\gamma\gamma$ optical depth (upper penal) and energy distribution (bottom penal) of $\gamma$-ray emission.  {\bf Upper penal:} the blue, yellow and red lines correspond to absorption due to TDE photons, host galaxy background light and EBL and CMB, respectively.
    	{\bf Bottom penal:} The predicted gamma-ray emission spectra during the outflow-cloud interactions. The blue, red and green lines present the spectrum for $\Gamma=1.5$, 1.7 and 2, respectively. The dotted and solid lines present the intrinsic and attenuated spectra. Also shown are the cumulative upper limits to the $\gamma$-ray flux observed by HAWC (dash-dot black line) and Fermi-LAT (dashed line). 
    	}
    	\label{GF}
    \end{figure}

    The $\gamma$-rays are emitted from clouds $\sim0.01$ pc away from the SMBH, so the absorption due to the TDE photon field relies on the emergent angle. We calculated the angle-averaged optical depth for the $\gamma$-rays (see the appendix J of \citealt{mou2020years1}). The optical depth due to TDE photons is found to be moderate for $\gamma$-rays of tens of GeV, as presented in Fig. \ref{GF}.
    
   Next, the absorption of the host galaxy's background light is considered. The photons are assumed to be isotropic, and we calculate $\gamma\gamma$ absorption as \cite{aharonian2004}. For host galaxy background photon fields, there was no more observation of the host galaxy 2MASS J20570298+1412165, except the infrared observation \citep{skrutskie2006}, namely J ($1.25\rm  \mu m$), H ($1.65\rm  \mu m$), and K ($2.16\rm  \mu m$). Thus we get the mean luminosity in J, H, K bands to be about $10^{42} \rm erg\ s^{-1}$. For the spectral profile we adopt the model in \cite{finke2010modeling} (the black line for redshift $z=0$ in Fig. 5 therein), and normalized to the infrared luminosity. The size of the host galaxy is typically of kpc scale. we find only mild absorption beyond TeV (see Fig. \ref{GF}).
   Furthermore, there would be significant absorption by the EBL and CMB. Considering the spectrum of EBL as model C in \cite{finke2010modeling} our calculation result for the EBL and CMB absorption is presented in Fig. \ref{GF}.
    Both intrinsic and attenuated spectra are plotted for comparison. The absorption significantly changes the emergent spectrum, mainly due to the EBL and CMB absorption. 

 Notice that we have considered also the absorption in the cloud since the high-energy $\gamma$-rays are produced in the cloud, but we find the attenuation in the cloud is negligible.  Firstly, the shocked cloud is optically thin to the high-energy gamma-rays for the electron-photon scattering between the thermal electron and $E_\gamma\sim\,\rm GeV$ high-energy photon, the optical depth for the GeV photon ${\tau _{e\gamma }} \simeq {r_{\rm c}}{n_{\rm c}}{\sigma _{\rm KN,GeV}} \simeq 6\times 10^{-4} r_{\rm c,14} n_{\rm c,10}$ with $\sigma _{\rm KN,GeV} \sim \sigma _{\rm T} (\varepsilon_{\gamma}/m_e c^2)^{-1} \simeq \sigma _{\rm T}/1000$ and the Bethe-Heitler process ${\tau _{\rm BH}} = {r_{\rm c}}{n_{\rm c}}{\sigma _{\rm BH}} \simeq 0.05{r_{\rm c,14}}{n_{\rm c,10}}$ for the parameters adopted in our model, i.e., cloud density $n_{\rm c,10}=n_{\rm c}/10^{10}\,\rm cm^{-3}$ and cloud radius $r_{\rm c,14}=r_{\rm c}/10^{14}\,\rm cm$. In addition to e-$\gamma$ scattering and Bethe-Heitler process for the high-energy gamma-rays, we evaluate the $\gamma\gamma$ absorption due to thermal radiations of clouds next. The shocked cloud emits through free-free radiation with a temperature $T_{\rm c} \approx {m_{\rm p}}V_{\rm c}^2/3k \approx 10^{7}\left( \frac{V_{\rm c}}{7\times 10^7 \, \rm cm/s} \right)^2\,\rm K$, and a luminosity for a single cloud ${L_{\rm X}} \simeq kT_{\rm c} * 4\pi r_{\rm c} ^3 n_{\rm c}/\max ({t_{\rm ff}},{t_{\rm sc}})\simeq 5\times 10^{37} \,\rm erg/s$, where $t_{\rm ff}\sim 2\times 10^{4} T_{\rm c,7}^{1/2}n_{\rm c,10}^{-1}\,\rm s$ is the cooling timescale of free-free radiation and $T_{\rm cloud}\sim 1.4 \times 10^{6}\,\rm s$ is the CS crossing timescale. The optical depth for the high-energy gamma-rays can be estimated as 
    \begin{equation}
        {\tau _{\gamma \gamma,c }} \sim {n_{\rm X}}{r_{\rm c}}{\sigma _{\gamma \gamma }} \sim 0.2{n_{\rm X}}{r_{\rm c}}{\sigma _{\rm T}} \sim {10^{ - 4}}n_{\rm X,7}r_{\rm c,14}\, 
    \end{equation}
    with the most optimistic cross section, where the number density can be written as ${n_{\rm X}} = \frac{{{L_{\rm X}}}}{{4\pi ck{T_{\rm c}}r_{\rm c}^2}} \simeq 1 \times {10^7}\,\rm c{m^{-3}}$ since the cloud is optically  moderately thin to its own radiations with the optical depth ${\tau _{e\gamma,{\rm c} }} \simeq {r_{\rm c}}{n_{\rm c}}{\sigma _{\rm T}} \simeq 0.6 r_{\rm c,14} n_{\rm c,10}$ (contributions of other clouds to number density can be easily neglected). Therefore, the opacity ($\tau _{e\gamma }, \tau _{\rm BH}, \tau _{\gamma \gamma }$) caused by the cloud to the high-energy gamma-rays can be neglected. In addition, considering the superposition of the free-free emission of $C_{\rm v} r_{\rm o}^2/r_{\rm c}^2 \sim 10^3$ clouds (total luminosity $\sim 10^3 L_{\rm X}\sim 5 \times 10^{40} \,\rm erg/s$), the corresponding total flux is quite low with a value $\sim 5 \times 10^{-15}\,\rm erg\, cm^{-2} s^{-1}$ at the keV energy band so that it is much lower than the observational upper limit of X-rays, even for the deep upper limit of $9 \times 10^{-14}\,\rm erg\, cm^{-2} s^{-1}$ (0.3-10 keV) given by \emph{XMM} observations in \cite{stein2021tidal}.

    
	\subsection{Other radiation}
	The BS accelerates both electrons and protons. The leptonic process of accelerated electrons also produces radiations. However, the ratio between the energy budgets of electrons and protons could be around $\sim 10^{-2}$ \citep{mou2020years1}, leading to a radiation luminosity by electrons is at most $\sim 10^{41}\rm erg\ s^{-1}$ for the fast cooling case, and the corresponding flux is quite low, $\sim 10^{-14}\rm erg\ cm^{-2}\ s^{-1}$ and can be neglected. 
	
	Moreover, the secondary electrons from $\gamma\gamma$ absorption may generate photons again via Inverse Compton scatterings, leading to the electromagnetic cascades. As shown in Fig.~\ref{GF}, only the EBL and CMB absorption is significant, and may results in electromagnetic cascades in the intergalactic medium. The deflection of electrons by the intergalactic magnetic field is expected to spread out the cascade emission, and contribute little to the observed flux.

	\section{Results compared with observations}\label{cwo}
	We summarize in Table \ref{123} the fiducial values for the model parameters used in the calculation. The neutrino luminosity is presented in Fig. \ref{NF}. According to equation \ref{n}, we get the expected event number of 0.1-1PeV neutrinos detected by IceCube. Without considering the partile escape from cloud, the expected number is $ 7.1\times10^{-3}$, $ 2.3\times10^{-3}$, $ 7.6\times10^{-4}$, and $3.8\times10^{-4}$ for $\Gamma=1.5$, 1.7, 1.9 and 2, respectively. However, if consider particle escape,  as described by equation \ref{eqescape}, the numbers change to $ 3.5\times10^{-3}$, $1.3\times10^{-3}$, $4\times10^{-4}$, and $1.9\times10^{-4}$, respectively.
	Thus, for the fiducial model parameters, the expected neutrino event number is somewhat lower than the expected neutrino number of $0.008 \lesssim N_\nu \lesssim 0.76$ for AT2019dsg \citep{stein2021tidal}.

 The interactions produce $\gamma$-rays from GeV to TeV energy bands. Considering the host galaxy distance of 230 Mpc, the photons above $\sim 100$ TeV cannot arrive at the Earth due to the absorptions by CMB and EBL. In addition, the strong absorption by the host galaxy photon field based on the infrared observations \citep{skrutskie2006} is moderate. Finally, the $pp$ processes produce the maximum $\gamma$-ray flux up to $\sim 10^{-13}\ \rm erg\ cm^{-2}\ s^{-1}$ for $\Gamma =1.5-2$ in the bands of 0.1 GeV - 1 TeV, which is lower than the present gamma-ray observational limits by Fermi/LAT and HAWC (see Fig. \ref{GF}).



	
	\section{Conclusion and discussion}

    In this work, we considered the high-speed TDE outflows colliding with the clouds to produce the high-energy neutrinos and gamma-rays, which can explain the sub-PeV neutrino event in AT2019dsg. The assumed outflow velocity is $V_{\rm o}\sim 0.07\rm c$ and the kinetic luminosity is $L_{\rm kin}\sim 10^{45}\rm erg~s^{-1}$. The outflow-cloud interactions will produce the BS ahead clouds. Particle acceleration is efficient in the BS and the pp process would contribute to the observational high energy neutrinos and gamma-rays. We assumed an escaping parameter of $F=0.5$, which means half of the accelerated protons escape from the BS region while the rest enter the dense cloud participating in pp collisions. Here we should point out that the escaping parameter of $F$ cannot be constrained well, leading to the $F$ values being quite uncertain, and the factor $F$ may be taken as a much smaller value in the more realistic cosmic-ray escape models. 
    
    For the fiducial model parameters, the expected neutrino event number would be relatively low compared to observations. In order to reach the observational neutrino number, one has to invoke some challenging model parameters. For instance, (1) considering a higher cloud density or a larger cloud size, inducing the escaping time $\tau_{\rm es}$ will be much higher than $t_{\rm pp,CS}$ (see equation \ref{eqescape}) and the interactions of protons which produce 0.1-1 PeV neutrinos becomes more efficient, the expected number of neutrinos would increase by a factor of $\sim 2$ (see Fig.~\ref{NF}); (2) the converting fraction of energy from the outflow to protons relies on the covering factor of BS $C_{\rm v}$, and shock acceleration efficiency $\eta$, so the expected number of neutrinos could increase by a factor of $\sim 10 (C_{\rm v} / 0.3)(\eta / 0.3)$ by taking a larger $C_{\rm v}$ and a larger $\eta$ than the fiducial values listed in Table.~\ref{123}. For sure, on the contrary, the expected neutrino number could be reduced if a lower cloud column density, smaller $C_{\rm v}$ and $\eta$, or a softer proton index is adopted. As a result, the predicted neutrino number in our model depends on the uncertainties of model parameters and in order to match the observations, some challenging values of parameters have been involved.

\begin{figure}
       \centering
       \includegraphics[scale=0.57]{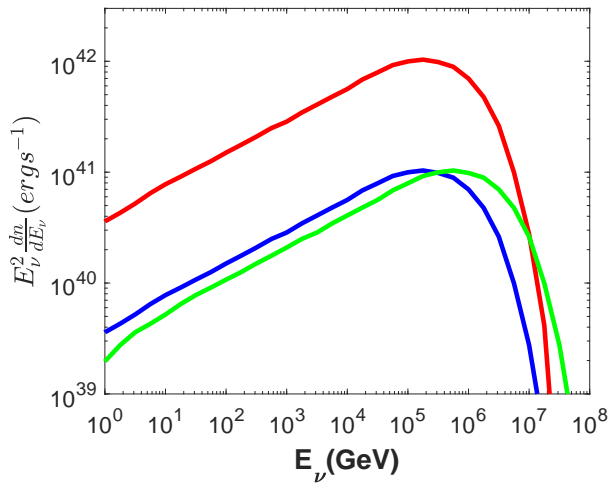}
       \caption{The neutrino luminosity as function of neutrino energy for different $L_{\rm kin}$ and $V_{\rm o}$, assuming $\Gamma =1.7$. The red line is the same as in Fig.3 with fiducial values. The blue line represents the case of $L_{\rm kin} = 10^{44}\ \rm erg/s$, resulting in the expected neutrino events of $1.3\times 10^{-5}$, about one order of magnitude lower than the fiducial value case. When the velocity of outflows is 0.12c, the maximum energy of accelerated protons reaches 180 PeV. The green line represents the case of $L_{\rm kin} = 10^{44}\ \rm erg/s$ and $V_{\rm o}= 0.12c$, leading to $E_{\rm max} = 180\ \rm PeV$ and the expected  neutrino events of $9\times10^{-6}$. 
       }
       \label{N44}
   \end{figure}

      In above calculations, we anchored two parameters of the outflows, the kinetic luminosity $L_{\rm kin}=10^{45} \rm erg~s^{-1}$ and the velocity of outflows $V_{\rm o}=0.07\rm c$. 
      Numberical simulations indicate that TDE can launch powerful wind with a kinetic luminosity of $10^{44-45}$ erg s$^{-1}$ \citep{curd2019grrmhd}, or even higher \citep{dai2018unified}. 
      AT2019dsg also exhibits radio flares, arising fifty days post burst and lasting for more than one year \citep{stein2021tidal,cendes2021radio}. Modeling the radio flare by the outflow-CNM (circumnuclear medium) model suggests that the averaged kinetic luminosity is $10^{43}$ erg s$^{-1}$ \citep{stein2021tidal} or even lower \citep{cendes2021radio}. 
      However, if the radio flare originates from outflow-cloud interaction which is the same scenario as our current model, the inferred kinetic luminosity may be in the order of $10^{44}$ erg s$^{-1}$ \citep{mou2021radio}. 
      Radio outflow and delayed neutrino may be explained by the same physical process.
      The detected neutrino number is linearly proportional to the kinetic luminosity. For the case of $L_{\rm kin}=10^{44}\rm erg ~s^{-1}$, the modeling neutrino luminosity is presented in Fig. \ref{N44}. 
      The expected neutrino number will be about one order of magnitude lower than the above values in the case of $L_{\rm kin}\sim 10^{45}\rm erg ~s^{-1}$.

      The velocity of the outflow is taken as $V_{\rm o}=0.07\rm c$, but this value is still uncertain, the radio observations suggested the outflow velocity in AT2019dsg is $V_{\rm o}=0.12\rm c$ \citep{stein2021tidal}, 0.07c \citep{cendes2021radio}
      or around 0.1c \citep{mou2021radio}. If the outflow velocity is higher, the maximum energy of accelerated protons in the BS will 
      be also higher accordingly. Then the peak neutrino luminosity will move to the higher energy ranges. If we integrate the neutrino number from $0.1-1$ PeV, the detected neutrino number would be different. For a comparison, we also plotted the neutrino luminosity versus energy in the case of $L_{\rm kin}=10^{44}~\rm{erg~s^{-1}}$, $V_{\rm o}=0.12\rm c$ in Fig. \ref{N44}. Since we only integrate the neutrino number in the range of $0.1-1$ PeV, if $V_{\rm o}$ is about 0.04c, and $E_{\rm p,max}\sim 20$ PeV, the neutrino SED will peak at $0.1-1$ PeV, and the expected number of neutrinos will increase by $50\%$.
      
     After the submission of this work, we notice recent reports on more neutrino events associated with the time-variable emission from accreting SMBHs \citep{van2021establishing,reusch2021candidate}, among which AT2019fdr is a TDE candidate in a Narrow-Line Seyfert 1 AGN in which the BLR clouds should exist. Moreover, the neutrino events lag the optical outbursts by half year to one year (\citealt{van2021establishing}), consistent with the assumption that clouds exist at the distance of $\sim 10^{-2}$ pc from the central BH if the outflow velocity is in the order of $10^9$ cm s$^{-1}$. The outflow--cloud interaction may also contribute to the high energy neutrino background \citep{abbasi2021icecube}.

\section*{Acknowledgments}
 We are grateful to the referee for the useful suggestions to improve the manuscript. This work is supported by the National Key Research and Development Program of China (Grants No. 2021YFA0718500, 2021YFA0718503), the NSFC (12133007, U1838103, 11622326, 11773008, 11833007, 11703022, 12003007, 11773003, and U1931201), the Fundamental Research Funds for the Central Universities (No. 2020kfyXJJS039, 2042021kf0224), and the China Manned Space Project (CMS-CSST-2021-B11).
\section*{Data Availability}
The data used in this paper were collected from the previous literatures. These data X-ray, gamma-ray and neutrino observations are public for all researchers. 
	
	
	
	\bibliographystyle{mnras}
	\bibliography{bib} 

	
	

\end{document}